\documentclass[preprint]{aastex}
\usepackage{psfig}
\slugcomment{KSUPT-02/1 \hspace{0.5truecm} July 2002}

\newcommand{\omegal}{\Omega_{\Lambda}}

\begin{document}

\title{Cosmological Constraints from Compact Radio Source Angular Size 
versus Redshift Data}

\author{Gang Chen and Bharat Ratra}

\affil{Department of Physics, Kansas State University, 116 Cardwell Hall, 
Manhattan, KS 66506.}

\begin{abstract}
We use the Gurvits, Kellermann, \& Frey compact radio source 
angular size versus redshift data to place constraints on cosmological
model parameters in models with and without a constant or time-variable
cosmological constant. The resulting constraints are consistent with
but weaker than those determined using current supernova apparent magnitude
versus redshift data.
\end{abstract}


\section{Introduction} 

Cosmological models now under consideration have a number of adjustable
parameters. A simple way to determine whether a model provides a 
useful approximation to reality is to use many different cosmological tests
to set constraints on cosmological-model-parameter values and to
check if these constraints are mutually consistent (see, e.g., Maor et al.
2002; Wasserman 2002).

During the past few years much attention has been focussed on the
Type Ia supernova apparent magnitude versus redshift test (see, e.g.,
Riess et al. 1998; Perlmutter et al. 1999; Podariu \& Ratra 2000;
Waga \& Frieman 2000; Gott et al. 2001; Leibundgut 2001).\footnote{
The proposed SNAP satellite should provide much tighter constraints
on cosmological parameters from this test (see, e.g., Podariu, Nugent,
\& Ratra 2001a; Weller \& Albrecht 2002; Wang \& Lovelace 2001; Gerke
\& Efstathiou 2002; Eriksson \& Amanullah 2002).}
This cosmological test indicates that the energy density of the current
universe is dominated by a cosmological constant, $\Lambda$, or a term 
in the material stress-energy tensor that only varies slowly with time and 
space and so behaves like $\Lambda$.

In conjunction with dynamical estimates which indicate a low non-relativistic
matter density parameter $\Omega_0$ (see, e.g., Peebles 1993), cosmic
microwave background anisotropy measurements also suggest the presence of
$\Lambda$ or a $\Lambda$-like term (see, e.g., Podariu et al. 2001b; Wang,
Tegmark, \& Zaldarriaga 2002; Baccigalupi et al. 2002; Durrer, 
Novosyadlyj, \& Apunevych 2001; Scott et al. 2002; Mason et al. 2002).

However, the observed rate of multiple images of radio sources or quasars,
produced by gravitational lensing by foreground galaxies, appears to favor
a smaller value for $\Lambda$ (see, e.g., Ratra \& Quillen 1992;
Helbig et al. 1999; Waga \& Frieman 2000; Ng \& Wiltshire 2001)
than is indicated by the observations mentioned above. It is therefore
of interest to examine the entrails of other cosmological tests.

In this paper we consider the redshift-angular size test, using the 
Gurvits, Kellermann, \& Frey (1999) compact radio source measurements.
The redshift-angular size relation is measured, for structures a few orders
of magnitude larger than those considered by Gurvits et al. (1999), by 
Buchalter et al. (1998) for quasars, and by Guerra, Daly, \& Wan (2000) 
for radio galaxies; we do not use these data sets in our analysis here. 
Vishwakarma (2001), Lima \& Alcaniz (2002), and references therein, use 
the Gurvits et al.(1999) data to set constraints on cosmological parameters; 
our results are consistent with, but as discussed next extend, their 
analyzes.\footnote{
The Gurvits et al. (1999) data augments that of Kellermann (1993). 
Stelmach (1994), Stepanas \& Saha (1995), Jackson \& Dodgson (1996),
and Kayser (1995) discuss the Kellermann (1993) data.}

Cosmological applications of the redshift-angular size test require 
knowledge of the linear size of the ``standard rod" used. Some earlier 
analyzes of the Gurvits et al. (1999) data appear to assume that this 
linear size will 
be determined using additional data, and so quote limits on cosmological
parameters (such as $\Omega_0$ or the cosmological constant density 
parameter $\omegal$) for a range of values of this linear size. Here
we note that it is best to treat this linear size as a 
``nuisance" parameter (for the cosmologically relevant part of this 
test), that is also determined by the redshift-angular size data, 
and so marginalize over it (using a prior to incorporate other information
about it, if needed).\footnote{
The situation here is similar to that for the redshift-magnitude test
(e.g., Riess et al. 1998; Perlmutter et al. 1999) where one must
marginalize over the magnitude of the standard candle used, treating
it as a nuisance parameter. In fact, Gurvits et al. (1999) determine the 
linear size from the redshift-angular size data by using the model of 
Gurvits (1994).}

In $\S$ 2 we summarize our computation. Results are 
presented and discussed in $\S$ 3. We conclude in $\S$ 4.

\section{Computation}

For our analyzes here we use the redshift-angular size data of Fig.
10 of Gurvits et al. (1999), which are binned redshift-angular size 
data derived from measurements of 145 sources. These measurements
are combined in twelve redshift bins, with about the same number of 
sources per bin, with the lowest and highest redshift bins centered 
at redshifts $z = 0.52$ and $z = 3.6$.

We consider two cosmological models as well as a currently popular
parametrization of dark energy. These are low-density cold dark
matter (CDM) dominated cases, consistent with current observational
indications. The first model is parametrized by two ``cosmological"
parameters, $\Omega_0$ and $\omegal$ (in addition to all the other usual
parameters). This model includes, as special cases, two ``one parameter"
models: the currently popular $\Lambda$CDM case with flat spatial
hypersurfaces and $\Lambda > 0$ (see, e.g., Peebles 1984; Efstathiou, 
Sutherland, \& Maddox 1990; Stompor, G\'orski, \& Banday 1995; Ratra 
et al. 1997; Sahni \& Starobinsky 2000) and a model with open spatial 
hypersurfaces and no $\Lambda$ (see, e.g., Gott 1982; Ratra \& Peebles 
1995; Cole et al. 1997). 

We also derive constraints on the parameters of a spatially-flat model
with a dark energy scalar field ($\phi$) with scalar field potential 
energy density $V(\phi)$ that at low $z$ is $\propto \phi^{-\alpha}$, 
$\alpha > 0$. The energy density of the scalar field decreases with
time, behaving like a time-variable $\Lambda$ (see, e.g., Peebles \& 
Ratra 1988; Ratra \& Peebles 1988; Steinhardt 1999; Brax, Martin,
\& Riazuelo 2000; Carroll 2001).

In linear perturbation theory, a scalar field is 
mathematically equivalent to a fluid with time-dependent 
equation of state parameter $w = p/\rho$ and speed of sound squared
$c_s^2 = dp/d\rho$, where $p$ is the pressure and $\rho$ the
energy density (see, e.g., Ratra 1991). The XCDM parametrization for 
dark energy approximates $w$ as a constant (see, e.g., Steinhardt 1999; 
Sahni \& Starobinsky 2000; Carroll 2001; Huterer \& Turner 2001), which 
is accurate in the radiation and matter dominated epochs but not in 
the current, dark energy scalar
field dominated epoch. Nevertheless the XCDM parametrization is 
recommended by its simplicity, so we also determine redshift-angular size 
constraints on its parameters. 

We want to determine how well the Gurvits et al. (1999) redshift-angular
size measurements distinguish between different 
cosmological-model-parameter values. To do this we pick one of the 
above models  or the XCDM parametrization, and a range of model-parameter 
values and compute the angular size distance $r (z)$ (the coordinate 
position for the object considered at redshift $z$ with the observer
at the origin) for a grid of model-parameter values that
span this range. An object of physical or proper length $l$ transverse to the
line of sight subtends an angle $\theta (z) = l (1 + z) / [a_0 r(z)]$, 
where $a_0$ is the current value of the scale factor (Peebles 1993, 
$\S$ 13). To determine the probability distribution of the
cosmological model parameters ($P$), we compute 
\begin{equation}
  \chi^2 (l, P) = \sum_{i = 1}^{12} \left[{{\theta(l, P, z_i) - 
      \theta_{\rm obs} (z_i)} \over \sigma(z_i)} \right]^2  ,
\end{equation}
where $\theta_{\rm obs} (z_i)$ and $\sigma (z_i)$ are the observed 
angles and errors for each of the twelve redshift bins centered 
at redshifts $z_i$ of the Gurvits et al. (1999) data. $P$ represents the
model parameters, for instance $\Omega_0$ and $\omegal$ in the general
two-dimensional constant $\Lambda$ case.  This representation (eq. [1]) 
is exact for the case where the correlated errors between redshift bins 
are negligible. 

With a uniform prior for the physical length $l$, the probability
distribution (likelihood) of the cosmological model parameters is 
\begin{equation}
  L(P) = \int dl \, e^{- \chi^2 (l, P)/2} ,
\end{equation}
where the integral is over a large enough range of $l$ to include
almost all of the probability. We typically integrate $l$ over the range 
1 $h^{-1}$ to 60 $h^{-1}$ pc,\footnote{
Here the Hubble constant $H_0 = 100 h$ km s$^{-1}$ Mpc$^{-1}$.} 
sampled at 60 points, although the range depends somewhat on the model 
under analysis. For the constant $\Lambda$ model we
compute over the ranges $ 0 \leq \Omega_0 \leq 1$ and $ -1 \leq	\omegal 
\leq 1$, both sampled at 201 points; for the scalar field dark energy model
we compute over the ranges $ 0.005 \leq \Omega_0 \leq 0.995$ and $ 0 \leq 
\alpha \leq 8$ sampled at 199 and 161 points, respectively; while for the 
XCDM parametrization we compute over the ranges $ 0 \leq \Omega_0 \leq 1$ 
and $ -1 \leq w \leq 0$, both sampled at 201 points. The confidence limits 
are computed from the distribution of eq. (2), and we typically show 1, 2, 
and 3 $\sigma$ confidence contours which correspond to enclosed probabilities 
of 68.27, 95.45, and 99.73 $\%$, respectively.

Since $\Omega_0$ is a positive quantity, we also consider the 
non-informative or logarithmic prior $p(\Omega_0) \propto 1/\Omega_0$ 
(Berger 1985; Gott et al. 2001) and compute confidence regions for the 
probability distribution $L(P)/\Omega_0$, where $L(P)$ is given in eq. (2).

\section{Results and Discussion} 

Figure 1 shows the Gurvits et al. (1999) redshift-angular size 
constraints on the general two-dimensional constant $\Lambda$ case. 
Apparently these have not previously been published. These constraints
are consistent with, but mostly not as constraining as those from Type Ia
supernova redshift-magnitude data, except at larger values of $\omegal$ and
$\Omega_0$ (see, e.g., Podariu \& Ratra 2000, Fig. 5).

Figure 2 shows the redshift-angular size constraints on the XCDM parameters.
Lima \& Alcaniz (2002, Fig. 2) show related constraints computed at 
fixed physical length $l$ for the same Gurvits et al. (1999) data.
While the shapes are similar, the Lima \& Alcaniz (2002) contours are much 
more constraining than those found here, largely because our procedure of
marginalizing over the physical length $l$ also accounts for the uncertainty
in the determination of $l$. Podariu \& Ratra (2000, Fig. 2) show 
corresponding constraints on the XCDM parameters from the Type Ia
supernova redshift-magnitude data, which are significantly more 
constraining that those shown in Fig. 2 here.

Figure 3 shows the constraints on the dark energy scalar field model 
with potential energy density $V(\phi) \propto \phi^{-\alpha}$, 
$\alpha > 0$ (Peebles \&  Ratra 1988). They are consistent with, but not as 
constraining as those from the Type Ia supernova redshift-magnitude data
(Podariu \& Ratra 2000; Waga \& Frieman 2000).

\section{Conclusion}

Constraints on cosmological model parameters derived from the redshift-angular
size compact radio source data of Gurvits et al. (1999) are consistent with 
but less constraining than those derived from the redshift-magnitude Type
Ia supernova data of Riess et al. (1998) and Perlmutter et al. (1999).

Higher quality redshift-angular size data will more significantly constrain
cosmological models, and in combination with high quality redshift-magnitude
data will provide a check of conventional general relativity on 
cosmological length scales.
   
\bigskip

We are indebted to L. Gurvits for providing the binned redshift-angular
size data. We acknowledge helpful discussions with J. Alcaniz, R. Daly,
J. Lima, and J. Peebles, and support from NSF CAREER grant AST-9875031.

%\clearpage

\begin{figure}[p]
\psfig{file=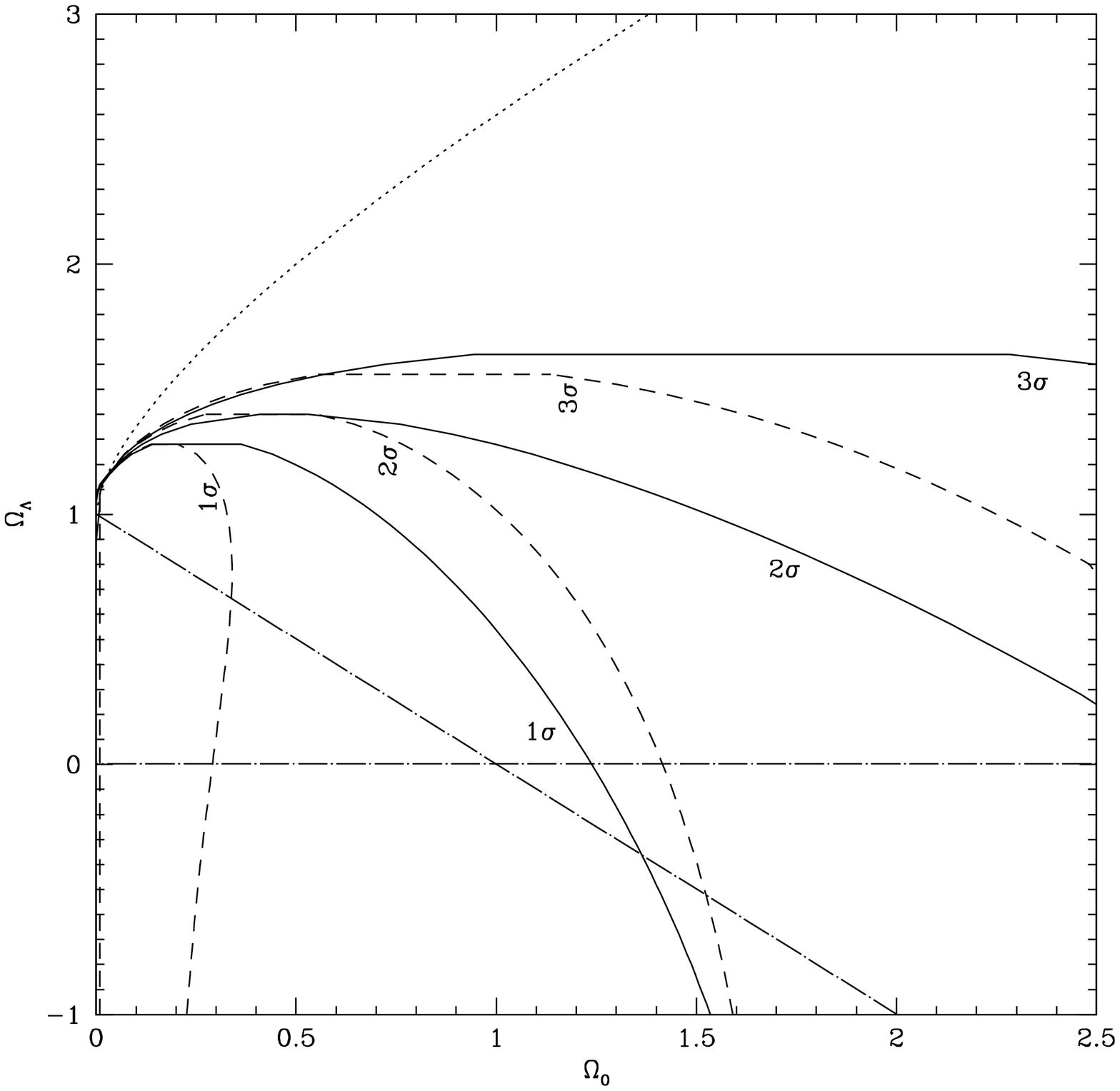,height=7.0in,width=6.7in,angle=0}
\caption{Contours of 1, 2, and 3 $\sigma$ confidence for the constant
$\Lambda$ model. Solid lines are contours computed using the uniform 
prior $p(\Omega_0) = 1$, while short dashed lines show the case using 
the logarithmic prior $p(\Omega_0) = 1/\Omega_0$ (with three contours
lying on each other at the left edge). The horizontal dot-dashed
line demarcates models with a vanishing cosmological constant, $\Lambda
= 0$, the dot-dashed line running from the point $\Omega_0 = 0$, 
$\omegal = 1$ to the point $\Omega_0 = 2$, $\omegal = -1$ indicates the 
spatially-flat $\Omega_0 + \omegal = 1$ case, and models lying in the 
upper left hand corner beyond the dotted line do not have a big bang.}
\end{figure}

\begin{figure}[p]
\psfig{file=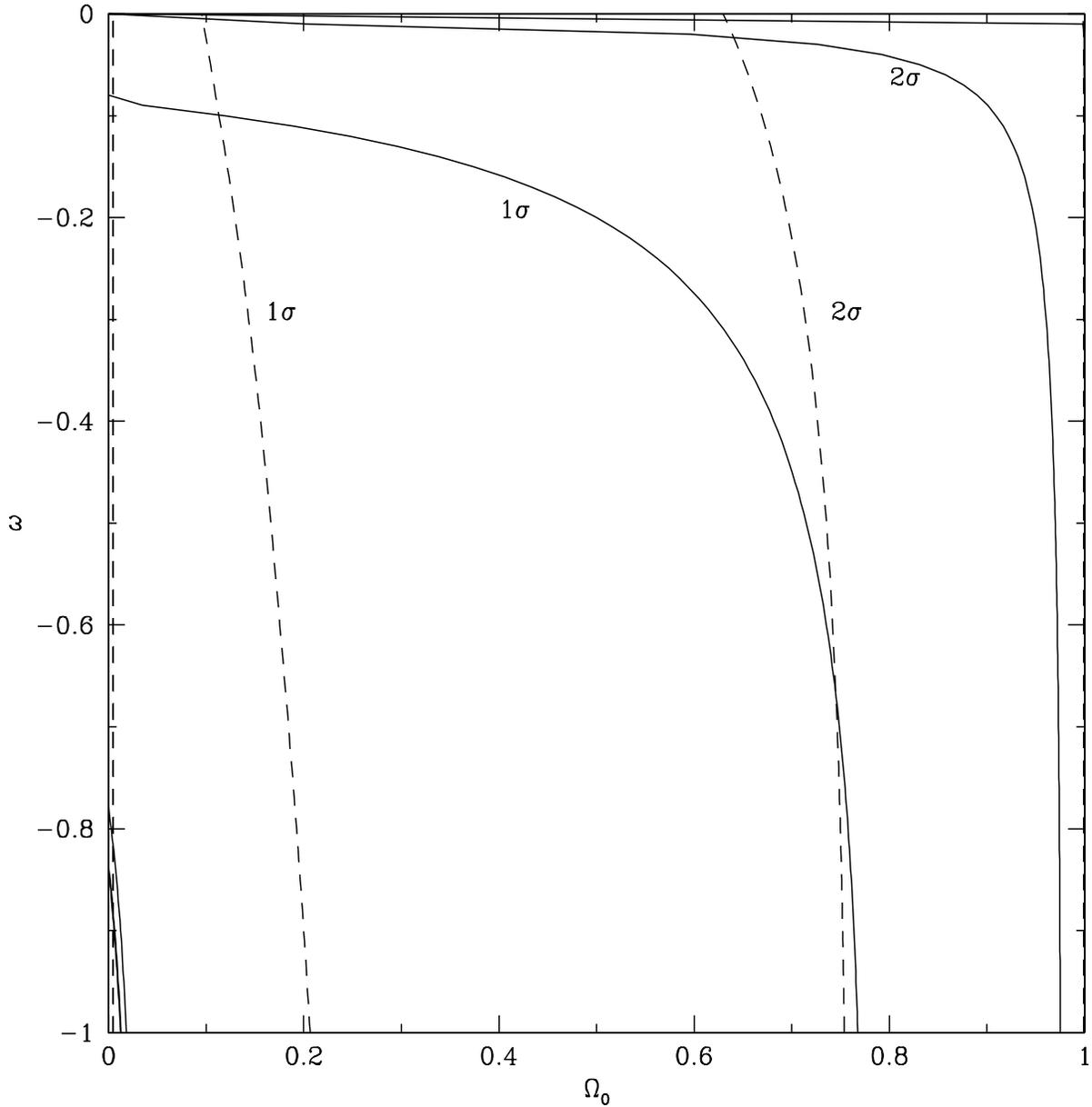,height=7.0in,width=6.7in,angle=0}
\caption{Contours of 1, 2, and 3 $\sigma$ confidence for the XCDM 
parametrization of dark energy, $p = w \rho$. Solid lines are contours 
computed using the uniform prior $p(\Omega_0) = 1$ (with three contours 
bunched together in the lower left corner), and short dashed lines show 
the case using the logarithmic prior $p(\Omega_0) = 1/\Omega_0$ (with 
three contours lying on each other at the left edge).}
\end{figure}

\begin{figure}[p]
\psfig{file=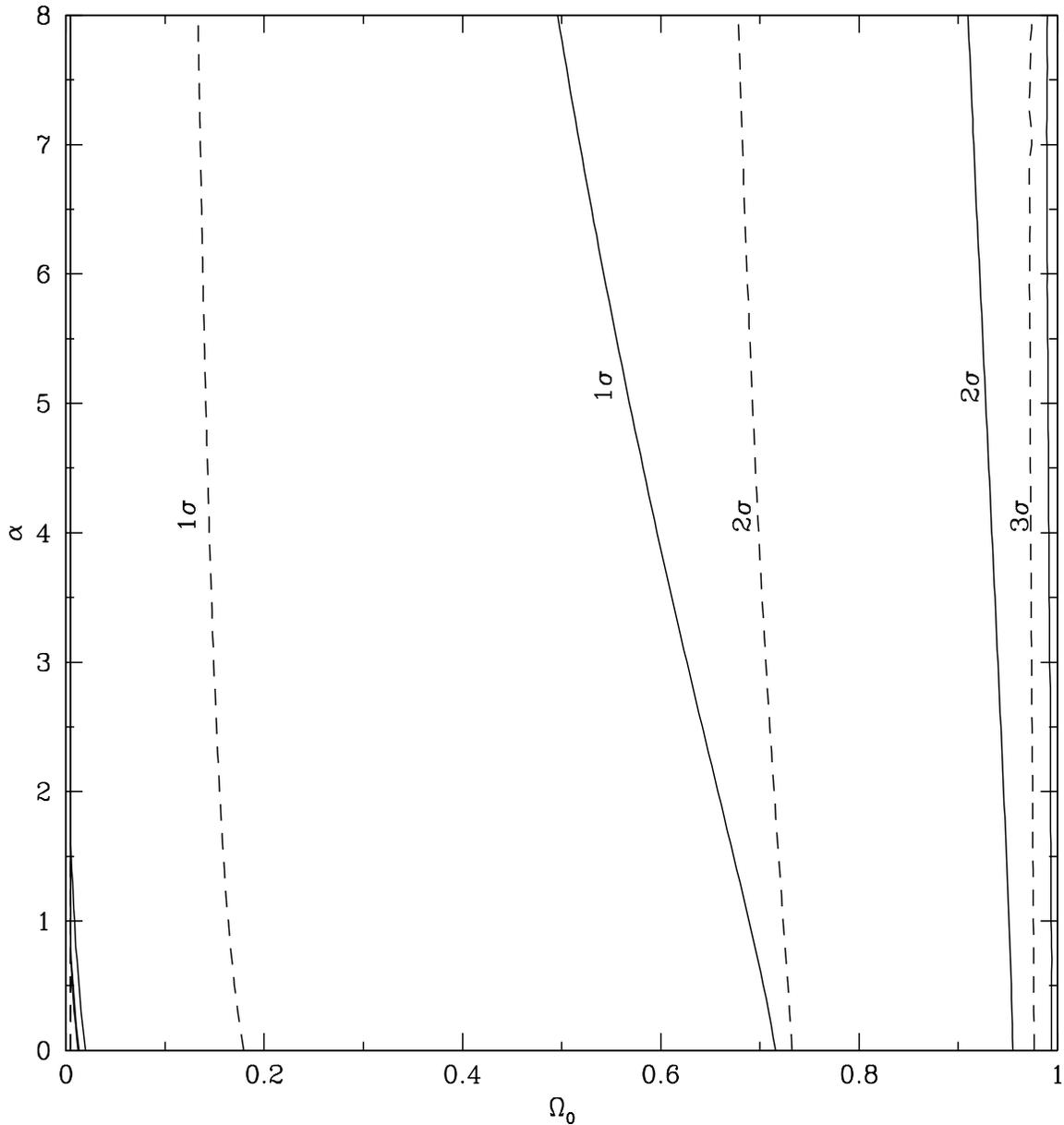,height=7.0in,width=6.7in,angle=0}
\caption{Contours of 1, 2, and 3 $\sigma$ confidence for the dark 
energy scalar field model with inverse power-law potential energy
density $V(\phi) \propto \phi^{-\alpha}$. Solid lines are contours 
computed using the uniform prior $p(\Omega_0) = 1$ (with three contours 
bunched together in the lower left corner), and short dashed lines show 
the case using the logarithmic prior $p(\Omega_0) = 1/\Omega_0$ (with 
three contours lying on each other at the left edge, at the boundary of 
parameter space).}
\end{figure}

\end{document}